\documentstyle[12pt]{article}
\newcommand{\eq}{\begin{equation}}
\newcommand{\en}{\end{equation}}
\newcommand{\eqa}{\begin{eqnarray}}
\newcommand{\ena}{\end{eqnarray}}
\newcommand{\eqs}{\begin{displaymath}}
\newcommand{\ens}{\end{displaymath}}
\newcommand{\eqas}{\begin{eqnarray*}}
\newcommand{\enas}{\end{eqnarray*}}

\advance\oddsidemargin by -\textwidth
\advance\oddsidemargin by -1in
\evensidemargin=\oddsidemargin
\begin{document}

$\mbox{ }$
\vspace{-3cm}
\begin{flushright}
\begin{tabular}{l}
{\bf KEK-TH-364 }\\
{\bf KEK preprint 93 }\\
{\bf UT-649 }\\
July 1993
\end{tabular}
\end{flushright}

\baselineskip18pt
\vspace{1cm}
\begin{center}
\Large
{\baselineskip26pt \bf String Field Theory\\
                       of Noncritical Strings}
\end{center}
\vspace{1cm}
\begin{center}
\large
$\mbox{{\sc Nobuyuki Ishibashi}}^{\star}~
\mbox{{\sc and Hikaru Kawai}}^{\dagger}$
\end{center}
\normalsize
\begin{center}
$\star$ {\it KEK Theory Group, Tsukuba, Ibaraki 305, Japan}\\
$\dagger$ {\it Department of Physics, University of Tokyo,}\\
{\it Bunkyo-ku, Tokyo 113, Japan}
\end{center}
\vspace{2cm}
\begin{center}
\normalsize
ABSTRACT
\end{center}
{\rightskip=2pc 
\leftskip=2pc 
\normalsize
We construct the Hamiltonian operator of the string field theory for $c=0$
string theory.
It describes how
strings evolve in the coordinate frame,
which is defined by using the geodesic distance
on the worldsheet.
The Hamiltonian consists of three-string interaction terms and a tadpole
term.
We show that one can derive the loop amplitudes of $c=0$ string theory
from this Hamiltonian.
\vglue 0.6cm}

\newpage

The matrix model provides the most powerful
technique for investigating noncritical string theories. The double scaling
limit \cite{DS} made it possible for us to discuss the sum of string
perturbation series. Therefore, for noncritical strings, the matrix
model gives us some clues about nonperturbative string effects. On the other
hand, for critical string theories, the string field theory \cite{SFT}
is supposed to be the most promising nonperturbative formulation of strings.
Since noncritical string theories can be considered as toy models of
critical strings, it is important to see what one can say about the string
field theory approach from the point of view of the matrix model formulation of
noncritical strings.

In this letter, we will construct a string field theory for $c=0$ string
theory. In order to do so, one should specify a way to cut string
worldsheets into fundamental pieces, i.e., propagator, vertex etc. This can
be done by introducing a time coordinate on the worldsheets.
Here we will use the time coordinate discussed in \cite{KKMW}
and its generalization. This time coordinate is a quite natural one in the
framework of the dynamical triangulation and the matrix model.
We will construct the string field Hamiltonian
operator corresponding to this time coordinate.
Our Hamiltonian includes only three
string vertices and a tadpole. In the latter part of this letter, we will
show that we can derive the string amplitudes using
this string field Hamiltonian. Remarkably, the Schwinger-Dyson equation of
this string field theory naturally yields the Virasoro constraints \cite{FKN}
deduced from the matrix model.

\vspace{1cm}
Let us first briefly recall the definition of the time coordinate in
\cite{KKMW}.
Suppose a disk with a metric.
One can define the time coordinate of a point as the
geodesic distance from the boundary loop.
Locally, the metric looks like
\eq
g_{00}=1,\;g_{01}=g_{10}=0,
\label{gauge}
\en
in such a coordinate frame.
This ADM-type gauge choice is
apparently more complicated compared to the conformal gauge and there is no
advantage in taking it in the continuum framework.
However, as was elucidated by the authors of \cite{KKMW}, such a definition of
time is rather natural in the framework of the dynamical triangulation.
They defined the operation called ``one-step deformation'' of a loop, which
exactly coincides with the discrete
evolution of the loop in the coordinate frame considered
here. In \cite{KKMW}, the authors showed that this ``one-step deformation''
has a well-defined continuum limit in the pure gravity case.

The disk can be considered as a closed string tadpole graph, where an incident
string disappears. First we restrict ourselves to this disk string
amplitude.
We would like to construct the string field ``Hamiltonian''
corresponding
to the time defined above and express the disk partition function
in terms of this Hamiltonian.
As is clear from the above paragraph, the easiest way to do so is to
construct a discrete Hamiltonian which describes the one-step
deformation on a triangulated disk and then take the continuum limit.
Let us look at the evolution of the
boundary loop ( or the incident string in the string terminology )
on a triangulated disk.
By just a short look at such an evolution,
one can easily see that
only the following two kinds of processes can occur:
\begin{enumerate}
\item The string splits into several strings,
\item The string disappears.
\end{enumerate}

We would like to construct a Hamiltonian ${\cal H}_{disk}$ representing
such processes in the
continuum limit, where $c=0$ string theory is realized.
In order to do so, it is convenient to consider the creation
and the annihilation operators of strings. Let $\Psi^{\dagger} (l)$
( $\Psi (l)$ ) be the creation ( annihilation ) operator which satisfies
\eq
\mbox{[} \Psi (l), \Psi^{\dagger }(l') \mbox{]}=\delta (l-l').
\label{comm}
\en
These operators create or annihilate a string with the length $l$.
To be precise, $\Psi^{\dagger }$ create a loop with one marked point and
$\Psi $ annihilate a loop with no marked point.
They act on the Hilbert space generated from the vacuum
$|0>$ and $<0|$:
\eqs
\Psi (l)|0>=<0|\Psi^{\dagger} (l)=0.
\ens
Then the two kinds of processes above would be expressed by the terms of
the form $(\Psi^{\dagger} )^n\Psi $ ( $n=0,1,2,...$ ) in ${\cal H}_{disk}$.
The disk amplitude corresponds to a process where the incident string
eventually disappears
after the time evolution dictated by the Hamiltonian.
The disk partition function will be expressed as
\eq
\lim_{D\rightarrow \infty}<0|e^{-D{\cal H}_{disk}}\Psi^{\dagger}(l)|0>.
\label{disk}
\en
Here $D$ is the Euclidean proper time, which we choose to be the time defined
above.

The Hamiltonian ${\cal H}_{disk}$ may be derived by constructing the full
discrete Hamiltonian through a brute force calculation
and taking the continuum limit. However,
without doing much calculation,
we can almost fix the form of ${\cal H}_{disk}$ by the following arguments.
The terms in
${\cal H}_{disk}$ are the ones which survive the continuum limit. Therefore
they should
have the appropriate scaling dimension.
Because of eq.(\ref{disk}), $\Psi^{\dagger}(l)$ should have the same scaling
dimension as that of the disk partition function. If one expresses the
dimension of $l$ by $L$,
$\mbox{[}\Psi^{\dagger}\mbox{]}=L^{-5/2}$ for $c=0$ \cite{KPZ}. Or in
other words, it should scale as $\epsilon ^{5/2}$ when the lattice spacing
$\epsilon$ approaches $0$ in the continuum limit. Eq.(\ref{comm}) implies
that $\mbox{[}\Psi \mbox{]}=L^{3/2}$. Since $\mbox{[}D \mbox{]}=L^{1/2}$
 \cite{KKMW},
${\cal H}_{disk}$ consists of the terms with the dimension $L^{-1/2}$.

The discrete Hamiltonian includes terms which express the
process where the string splits into $n$ strings, i.e.,  $n+1$-
string interaction term, with $n=1,2,3,...$.
However,
by a brief examination of the one-step deformation, one can see that the
most dominant contribution of the $n+1$-
string interaction term in the continuum limit $\epsilon\rightarrow 0$ is
of the form
\eqs
\epsilon^{\frac{n-1}{2}}\int \int dl_1\cdots dl_n
                \Psi^{\dagger}(l_1)\cdots \Psi^{\dagger}(l_n)
                \Psi (l_1+\cdots l_n)(l_1+\cdots l_n)^{n-1}.
\ens
Therefore the only terms which survive the continuum limit and contribute to
${\cal H}_{disk}$ with the dimension $L^{-1/2}$ are the ones with $n=1,2$.
Adding the
tadpole term, by which the string disappears,
the most general form of such a Hamiltonian is
\eqa
{\cal H}_{disk}
&=&
 \int_0^{\infty}dl_1\int_0^{\infty}dl_2
  \Psi^{\dagger}(l_1)\Psi^{\dagger}(l_2)\Psi (l_1+l_2)(l_1+l_2)
\nonumber
\\
& &
+\int_0^{\infty}dl_1\int_0^{\infty}dl_2
  \Psi^{\dagger}(l_1)K(l_1,\;l_2)\Psi (l_2)
\nonumber
\\
& &
+\int _0^{\infty}dl\rho (l)\Psi (l).
\label{hamd}
\ena
We can always normalize ${\cal H}_{disk}$ so that
the coefficient of the first term is
$1$,
by rescaling the definition of the time $D$. The second term is the string
kinetic term and the third term is the tadpole term.

It is remarkable that the above dimensional analysis is powerful enough to
specify the form
of the Hamiltonian as eq.(\ref{hamd}). Although there may be infinitely many
terms in the discrete Hamiltonian, only three of them survive the continuum
limit. In order to fix ${\cal H}_{disk}$ completely, one should obtain the
kernel $K(l_1,l_2)$ and $\rho (l)$.
Here we would like to claim that the kinetic term is absent and the
Hamiltonian takes the pregeometric form
\eqa
{\cal H}_{disk}
&=&
 \int_0^{\infty}dl_1\int_0^{\infty}dl_2
  \Psi^{\dagger}(l_1)\Psi^{\dagger}(l_2)\Psi (l_1+l_2)(l_1+l_2)
\nonumber
\\
& &
+\int _0^{\infty}dl\rho (l)\Psi (l).
\label{hamp}
\ena
Here we will not prove $K(l_1,l_2)=0$ but rather proceed assuming
$K(l_1,l_2)=0$.
The validity of this assumption will be checked later by
showing
that the pregeometric Hamiltonian can reproduce loop amplitudes of $c=0$
string theory.

$\rho (l)$ can be derived by imposing the condition that
eq.(\ref{disk}) gives the disk partition function of the pure
gravity. Let us derive a Schwinger-Dyson equation from eq.(\ref{disk}).
The
existence of the large $D$ limit in eq.(\ref{disk}) implies the following
equation:
\eq
\lim_{D\rightarrow \infty}\frac{\partial}{\partial D}
      <0|e^{-D{\cal H}_{disk}}\Psi^{\dagger}(l)|0>=0.
\label{partD}
\en
This equation means that in the large $D$ limit,
$<0|e^{-D{\cal H}_{disk}}\Psi^{\dagger}(l)|0>$ does not evolve any more.
Hence eq.(\ref{partD}) corresponds to
the Wheeler deWitt equation for the wave function
$f(l)=\lim_{D\rightarrow \infty}<0|e^{-D{\cal H}_{disk}}\Psi^{\dagger}(l)|0>$
in the gauge
eq.(\ref{gauge}).
Since ${\cal H}_{disk}|0>=0$, eq.(\ref{partD}) becomes
\eq
\lim_{D\rightarrow \infty}
      <0|e^{-D{\cal H}_{disk}}
      \mbox{[}{\cal H}_{disk},\Psi^{\dagger}(l)\mbox{]}|0>=0.
\label{partd}
\en
Using the fact that strings can only split and never merge in ${\cal
H}_{disk}$,
it is easy to prove the factorization
\eqas
& &
\lim_{D\rightarrow \infty}
<0|e^{-D{\cal H}_{disk}}\Psi^{\dagger}(l_1)\Psi^{\dagger}(l_2)|0>
\\
& &
\hspace{3cm}
=
\lim_{D\rightarrow \infty}<0|e^{-D{\cal H}_{disk}}\Psi^{\dagger}(l_1)|0>
\lim_{D\rightarrow \infty}<0|e^{-D{\cal H}_{disk}}\Psi^{\dagger}(l_2)|0>.
\enas
Thus
one obtains the Schwinger-Dyson equation ( or the Wheeler deWitt equation ) for
$f(l)$
from eq.(\ref{partd}) as
\eq
l\int_0^{l}dl_1f(l_1)f(l-l_1)
+\rho (l)=0.
\label{SDd}
\en
Notice that this Wheeler deWitt equation is nonlinear and inhomogeneous
contrary to the conformal gauge Wheeler deWitt equation \cite{MSS}. In the
conformal gauge, the uniformization theorem for Riemann surfaces
implies that the incident string
never splits when one considers the time evolution of the boundary loop on
the disk. However, in the gauge eq.(\ref{gauge}), it seems that
splitting and disappearing are
the main part of the time evolution.

Laplace transforming eq.(\ref{SDd}), we obtain
\eqs
-\partial_{\zeta}(\tilde{f}(\zeta )^2)+\tilde{\rho }(\zeta )=0,
\ens
where
\eqs
\tilde{f}(\zeta )=\int_0^{\infty }dle^{-\zeta l}f(l),\;
\tilde{\rho }(\zeta )=\int_0^{\infty }dle^{-\zeta l}\rho (l).
\ens
This equation is easily solved as
\eqs
\tilde{f}(\zeta )=\sqrt{\int^{\zeta}d\zeta '\rho (\zeta ')}.
\ens
The result $\tilde{f}(\zeta )
=(\zeta -\frac{1}{2}\sqrt{t})\sqrt{\zeta +\sqrt{t}}$ of the matrix model
\cite{BIPZ} with $t$ as the cosmological constant,
completely fixes $\tilde{\rho}$:
\eqs
\tilde{\rho }(\zeta )=3\zeta ^2 -\frac{3}{4}t,
\ens
or
\eqs
\rho (l)=3\delta ''(l)-\frac{3}{4}t\delta (l).
\ens
The fact that $\rho (l)$
has its support only at $l=0$ is quite natural because it means that only
strings with zero length can disappear.

Here as a check of the validity of our disk Hamiltonian ${\cal H}_{disk}$,
let us derive the results in \cite{KKMW}
using ${\cal H}_{disk}$. In that paper, they
concentrated on so to speak {\it inclusive} processes in which they kept
track of only {\it one} of the loops to which the incident loop split after
the time evolution. Given the Hamiltonian eq.(\ref{hamp}) describing the whole
process, it is easy to see the Hamiltonian $H$ for the {\it inclusive} process
should be
\eq
H=2\int_0^{\infty}dl_1\int_0^{\infty}dl_2
  f(l_1)\Psi^{\dagger}(l_2)\Psi (l_1+l_2)(l_1+l_2).
\en
The ``proper time
evolution kernel''  $\tilde{N}(\zeta , \zeta ';D;t)$ defined in \cite{KKMW}
can be expressed as
\eq
\tilde{N}(\zeta , \zeta ';D;t)
=
\int_0^{\infty }dl\int_0^{\infty }dl'e^{-\zeta l-\zeta 'l'}\frac{l'}{l}
<0|\Psi (l')e^{-DH}\Psi^{\dagger}(l)|0>.
\en
Notice the factor $\frac{l'}{l}$ which is necessary because the exit loop is
marked and the entrance one is not in $\tilde{N}(\zeta , \zeta ';D;t)$. From
this expression, we obtain a formula
\eq
\frac{\partial}{\partial D}\tilde{N}(\zeta , \zeta ';D;t)
=
2\tilde{f}(\zeta )\frac{\partial}{\partial \zeta}\tilde{N}(\zeta , \zeta
';D;t).
\en
This coincides with the equation for $\tilde{N}(\zeta , \zeta ';D;t)$ up to
a rescaling of $D$.

So far we have been considering only the disk amplitude where one incident
string disappears after the time evolution. We will now go on to the amplitudes
corresponding to connected surfaces with many boundaries.
Generalizing the disk case,
the time coordinate of a point can be defined as the geodesic distance of the
point from the union of the boundary loops.
With more than one incident loops,
one should take into account the process in which two strings merge in addition
to the two kinds of processes we have considered so far. The
dimensional analysis makes it possible to deduce that the total Hamiltonian
including such terms becomes
\eqa
{\cal H}
&=&
\int_0^{\infty}dl_1\int_0^{\infty}dl_2
  \Psi^{\dagger}(l_1)\Psi^{\dagger}(l_2)\Psi (l_1+l_2)(l_1+l_2)
\nonumber
\\
& &
  +g\int_0^{\infty}dl_1\int_0^{\infty}dl_2
  \Psi^{\dagger}(l_1+l_2)\Psi (l_1)\Psi (l_2)l_1l_2
\nonumber
\\
& &
+\int _0^{\infty}dl\rho (l)\Psi (l).
\ena
Here $g$ is the string coupling constant whose dimension is
$\mbox{[}g\mbox{]}=L^{-5}$ as was derived in \cite{DS}.
The partition function for the surfaces with $n$ boundaries should be expressed
as
\eq
\lim_{D\rightarrow \infty}
<0|e^{-D{\cal H}}\Psi^{\dagger }(l_1)\cdots \Psi^{\dagger }(l_n)|0>.
\label{nloop}
\en
Because of the
merging process, this expression automatically includes the
contributions from the surfaces with more than one handles.
Expanding perturbatively in terms of $g$, the contribution from the
connected surfaces
with $h$ handles and $b$ boundaries is proportional to $g^{-1+h+b}$.

Thus we have derived the string field Hamiltonian from
the dimensional analysis and the assumptions such as the absence of the
kinetic term.
It consists of three string vertices and a tadpole. It is reminiscent of
the light-cone gauge string field theory of the critical strings.
In the rest of this letter, we will show that this simple
Hamiltonian can reproduce the loop ( or string ) amplitudes
of $c=0$ string theory.

\vspace{1cm}
Let us derive the macroscopic loop amplitudes \cite{MSS} ( or multi-string
amplitudes ) from ${\cal H}$. This may be done by calculating eq.(\ref{nloop})
perturbatively in $g$.
Here, we will rather derive the string field Schwinger-Dyson equations
for the multi-string amplitudes. Imposing a physical boundary condition,
it is possible to prove that the
Schwinger-Dyson equations have unique solutions, if any. We will show that
the loop amplitudes of $c=0$ string theory provide the unique solutions to
the Schwinger-Dyson equations by demonstrating that the Schwinger-Dyson
equations naturally yield the Virasoro constraints of the matrix model.

In order to
do so, we construct a Schwinger-Dyson equation for the generating
functional of the loops, i.e.,
\eq
Z(J)=
\lim_{D\rightarrow \infty}
<0|e^{-D{\cal H}}e^{\int dlJ(l)\Psi^{\dagger }(l)}|0>.
\en
The method to obtain a Schwinger-Dyson equation is the same as in
eq.(\ref{partD}). The existence of the large $D$ limit implies
\eqs
\lim_{D\rightarrow \infty}
\frac{\partial }{\partial D}
<0|e^{-D{\cal H}}e^{\int dlJ(l)\Psi^{\dagger }(l)}|0>
=0.
\ens
{}From the view point of two dimensional gravity, this equation again has
the meaning of the Wheeler deWitt equation.
Using
\eqs
<0|e^{-D{\cal H}}e^{\int dlJ(l)\Psi^{\dagger }(l)}|0>=
<0|e^{-D{\cal H}^{\prime }}|0>,
\ens
where
\eqs
{\cal H}^{\prime }
=
e^{-\int dlJ(l)\Psi^{\dagger }(l)}
{\cal H}
e^{\int dlJ(l)\Psi^{\dagger }(l)},
\ens
the Schwinger-Dyson equation becomes
\eqa
& &
\int_0^{\infty }dl_1\int_0^{\infty }dl_2J(l_1+l_2)(l_1+l_2)
\frac{\delta}{\delta J(l_1)}\frac{\delta}{\delta J(l_2)}Z(J)
\nonumber
\\
& &
+g\int_0^{\infty }dl_1\int_0^{\infty }dl_2J(l_1)J(l_2)l_1l_2
\frac{\delta}{\delta J(l_1+l_2)}Z(J)
\nonumber
\\
& &
+\int_0^{\infty }dl\rho (l)J(l)Z(J)=0.
\label{eqZ}
\ena
For later use, we will
transform
eq.(\ref{eqZ}) into an equation for the generating functional $\ln Z(J)$
of the connected amplitudes:
\eqa
& &
\int_0^{\infty }dlJ(l)\{
l\int_0^ldl'J(l')
\mbox{[}
\frac{\delta^2\ln Z(J)}{\delta J(l')\delta J(l-l')}
+\frac{\delta \ln Z(J)}{\delta J(l)}\frac{\delta \ln Z(J)}{\delta J(l-l')}
\mbox{]}
\nonumber
\\
& &
\hspace{2cm}
+gl\int_0^{\infty }dl'J(l')l'
\frac{\delta \ln Z(J)}{\delta J(l+l')}
\nonumber
\\
& &
\hspace{2cm}
+\rho (l)\} =0.
\label{eqc}
\ena
We regard $\ln Z(J)$ as the generating functional of the connected correlation
functions of loop operators $w(l)$, i.e.,
\eq
\ln Z(J)=<e^{\int dlJ(l)w(l)}>.
\en
The connected loop amplitude $<w(l_1)\cdots w(l_n)>$ can be expressed as
\eq
<w(l_1)\cdots w(l_n)>=
\frac{\delta^n \ln Z(J)}{\delta J(l_1)\cdots \delta J(l_n)}|_{J=0}.
\en

Eq.(\ref{eqc}) can be regarded as the
generating functional of Schwinger-Dyson equations.
By differentiating this equation several times and putting $J=0$ after that,
one obtains an equation for connected loop amplitudes.
The connected loop amplitudes can be expanded in terms
of $g$ as
\eq
<w(l_1)\cdots w(l_n)>=\sum_{h=0}^{\infty }g^{n-1+h}<w(l_1)\cdots w(l_n)>_h.
\label{pert}
\en
Here $<w(l_1)\cdots w(l_n)>_h$ is the contribution from the surfaces with
$h$ handles. Substituting eq.(\ref{pert}) into eq.(\ref{eqc}), one obtains
equations of the form
\eqa
& &
\sum_{k=1}^n\partial_{\zeta_k}(<\tilde{w}(\zeta_1)\cdots \tilde{w}(\zeta_n)>_h
                   <\tilde{w}(\zeta_k)>_0)
\nonumber
\\
& &
\hspace{1cm}
=\mbox{sum~of~the~terms~made~from~}
<\tilde{w}(\zeta_1)\cdots \tilde{w}(\zeta_m)>_{h'}
\nonumber
\\
& &
\hspace{3cm}
(m<n,\;h=h'\mbox{~or~}m=n+1,\;h'=h-1),
\label{eqf}
\ena
where $<\tilde{w}(\zeta_1)\cdots \tilde{w}(\zeta_n)>_h$ is the Laplace
transform
\eqs
<\tilde{w}(\zeta_1)\cdots \tilde{w}(\zeta_n)>_h
=
\int_0^\infty dl_1e^{-\zeta_1l_1}\cdots \int_0^\infty dl_ne^{\zeta_nl_n}
<w(l_1)\cdots w(l_n)>_h.
\ens
Therefore, in principle, it is possible to solve all these equations
inductively
starting from the amplitudes with fewer loops and handles.

In order to solve each of these
Schwinger-Dyson-Wheeler-deWitt equations, we should impose an appropriate
boundary condition. Here we will require that $<w(l_1)\cdots w(l_n)>$
vanishes when any of $l_i$ goes to infinity. Considering the loop amplitudes
as the wave function of two dimensional quantum gravity corresponding to a
multi-loop ( or multi-universe ) state, it is natural to impose such a
condition \cite{MSS}. This condition implies that
$<\tilde{w}(\zeta_1)\cdots \tilde{w}(\zeta_n)>_h$ does not have any
singularities when the real parts of all the $\zeta_i$ are positive.
It is possible to
show that if the equation of the form eq.(\ref{eqf}) has a solution
$<\tilde{w}(\zeta_1)\cdots \tilde{w}(\zeta_n)>_h$ satisfying such a boundary
condition, the solution is unique.

Thus eq.(\ref{eqc}) has a unique solution, if any.
We would like to show that the loop amplitudes of $c=0$ string theory provide
 the
unique solution to eq.(\ref{eqc}).
Notice that if $Z(J)$ satisfies
\eqa
& &
l\int_0^ldl'J(l)
\mbox{[}
\frac{\delta^2\ln Z(J)}{\delta J(l')\delta J(l-l')}
+\frac{\delta \ln Z(J)}{\delta J(l)}\frac{\delta \ln Z(J)}{\delta J(l-l')}
\mbox{]}
\nonumber
\\
& &
+gl\int_0^{\infty }dl'J(l')l'
\frac{\delta \ln Z(J)}{\delta J(l+l')}
\nonumber
\\
& &
+\rho (l)=0,
\label{sdeq}
\ena
it is a solution of eq.(\ref{eqc}). Of course, eq.(\ref{sdeq}) is not totally
equivalent to eq.(\ref{eqc}). Eq.(\ref{eqc}) is obtained if one symmetrizes
eq.(\ref{sdeq}) over all the incident loops. However, because of the uniqueness
of the solution, if eq.(\ref{sdeq}) has a solution satisfying the boundary
condition, it is also the unique solution of eq.({\ref{eqc}). Eq.(\ref{sdeq})
looks quite similar to the loop equation of the matrix model.
Indeed, in the following, we will derive the
Virasoro constraints \cite{FKN} of $c=0$ string theory from this equation.

In \cite{FKN}, the authors transform the loop equation into the relations
between the correlation functions of the local operators ${\cal O}_n$,
which appear in the expansion of the loop operator as
\eq
w(l)=g\sum_{n\geq 0}\frac{l^{n+1/2}}{\Gamma (n+\frac{3}{2})}{\cal O}_n.
\label{local}
\en
The factor $g$ on the right hand side is put so that the insertions of
local operators do not change the order in $g$.
In order to drive equations for amplitudes with insertions of such local
operators, we should choose the source $J(l)$ so that
\eq
\int_0^{\infty}dlJ(l)l^{n+\frac{1}{2}}=g^{-1}\Gamma (n+\frac{3}{2})\mu_n.
\label{source}
\en
Then the generating
functional $\ln Z(J)$ can be considered as the generating functional
of connected correlation functions of the local operators:
\eqs
\ln Z(J)=<e^{\sum \mu_n{\cal O}_n}>.
\ens

Therefore substituting eq.(\ref{source}) into eq.(\ref{sdeq}), we can obtain
relations between the correlation functions of the local operators. One thing
one should notice in doing so is that the loop operator $w(l)$ cannot
always be expanded as eq.(\ref{local}).
Since the amplitudes for the disk and the cylinder have a special part which
cannot be written as eq.((\ref{local}),
we have the following ansatz for the solution of
eq.(\ref{sdeq}):
\eqas
\frac{\delta \ln Z(J)}{\delta J(l)}
=
\frac{l^{-\frac{5}{2}}}{\Gamma (-\frac{3}{2})}
-\frac{3t}{8}\frac{l^{-\frac{1}{2}}}{\Gamma (\frac{1}{2})}
+\frac{g}{2\pi}\int_0^{\infty}dl'J(l')\frac{\sqrt{ll'}}{l+l'}
+g\sum_{n\geq 0}\frac{l^{n+1/2}}{\Gamma (n+\frac{3}{2})}
\frac{\delta \ln Z(J)}{\delta \mu_n}.
\enas
Substituting this, the Laplace transform of the Schwinger-Dyson equation
eq.(\ref{sdeq}) becomes
\eqa
& &
-\partial_{\zeta}
\mbox{[}
(\zeta^{\frac{3}{2}}-\frac{3t}{8}\zeta^{-\frac{1}{2}}
+g\sum_{n\geq 0}\zeta^{-n-\frac{3}{2}}\frac{\delta \ln Z(J)}{\delta \mu_n})^2
+\frac{g}{16}\frac{1}{\zeta^2}
\nonumber
\\
& &
\hspace{1cm}
+g^2\sum_{n,m\geq 0}\zeta^{-n-m-3}
\frac{\delta^2 \ln Z(J)}{\delta \mu_n\delta \mu_m}
\nonumber
\\
& &
\hspace{1cm}
+(\frac{\mu_0^2}{16}-\frac{3t}{16}\mu_0)\frac{1}{\zeta}
+g\sum_{n\geq 0}\frac{\delta \ln Z(J)}{\delta \mu_n}
\sum_{m=0}^{n+1}\zeta^{-n+m-2}(m+\frac{1}{2})\mu_m
\mbox{]}
\nonumber
\\
& &
+3\zeta^2-\frac{3}{4}t=0.
\label{SDL}
\ena
Since the right hand side of eq.(\ref{SDL}) is in the form of the Laurent
expansion in terms of $\zeta^{-1}$, each coefficient of the expansion should
vanish. Thus we obtain the following infinite number of equations.
\eqa
& &
2\frac{\partial Z}{\partial \mu_1}
=
-\frac{1}{g}(\frac{3t}{8}-\frac{\mu_0}{2})^2Z
-\sum_{n=1}^{\infty}(n+\frac{1}{2})\mu_n\frac{\partial Z}{\partial \mu_{n-1}},
\\
& &
2(\frac{\partial Z}{\partial \mu_2}
-\frac{3t}{8}\frac{\partial Z}{\partial \mu_0})
=
-\frac{1}{16}Z
-\sum_{n=0}^{\infty}(n+\frac{1}{2})\mu_n\frac{\partial Z}{\partial \mu_{n}},
\\
& &
2(\frac{\partial Z}{\partial \mu_{p+3}}
-\frac{3t}{8}\frac{\partial Z}{\partial \mu_{p+1}})
=
-g\sum_{n=0}^p\frac{\partial^2 Z}{\partial \mu_n\mu_{p-n}}
-\sum_{n=0}^{\infty}(n+\frac{1}{2})\mu_n\frac{\partial Z}{\partial \mu_{n+p+1}}
\\
& &
\hspace{7cm}(p\geq 0).
\nonumber
\ena
These equations coincide with the Virasoro constraints in \cite{FKN} up to
some rescalings of parameters. Therefore the loop amplitudes of two dimensional
gravity provide the unique solution of eq.(\ref{sdeq}).

\vspace{1cm}
To sum up, we have derived all the string
( or loop ) amplitudes in $c=0$ string theory ( or pure gravity ) from
our string field Hamiltonian. It is
remarkable that the {\it string field} Schwinger-Dyson equation yields the
Virasoro constraints which are equivalent to the
{\it matrix model} Schwinger-Dyson
equation.
The reason is simple if one considers at the discretized level.
The matrix model Schwinger-Dyson equation
describes how a loop amplitude behaves when one deforms the loop at one
link on the loop.
The string field Schwinger-Dyson equation describes how
a loop amplitude behaves when one deforms all the incident loops
one-step forward in the
sense of \cite{KKMW}. It is intuitively
clear that in the continuum limit
the latter one is obtained if we integrate the former one around the loop
 and symmetrize over the incident loops.
The advantage of our string field Hamiltonian is that it is related to the
time coordinate inherent in the matrix model formulation of noncritical string
theories. This makes the string field Schwinger-Dyson equation be related to
the matrix model Schwinger-Dyson equation. Moreover, the physical
meaning of the Virasoro constraints is now clear. They are the Wheeler deWitt
equations of two dimensional gravity in the ADM-type gauge eq.(\ref{gauge}).

We have derived the string field Hamiltonian for $c=0$ string theory
from which one can
reproduce all the known results of $c=0$ string theory.
It implies that our string field theory is good enough to cover the
moduli spaces of Riemann surfaces of all genera only once, with the
three string vertices and the tadpole term.
We expect that
the same method can be applied to the critical strings. In order do so,
one should study the string theory with some matter fields on the worldsheet.
We will report on this problem elsewhere.

\section*{Acknowledgements}
We would like to thank N.Kawamoto, T.Mogami, Y.Okamoto, Y.Watabiki
and T.Yukawa for useful discussions and comments.

\end{document}